\documentclass[pra,aps,twocolumn,10pt,showpacs,groupedaddress,superscriptaddress,floatfix,notitlepage]{revtex4-2}
\usepackage{amsmath,amsfonts,amssymb,graphics,graphicx,epsfig,color,times}
\usepackage[utf8x]{inputenc}
\usepackage{color}
\usepackage{bbm, dsfont} 
\usepackage{subfigure}
\usepackage{hyperref}
\usepackage{mathrsfs}
\usepackage{verbatim}

\begin{document}
\title{Chiral-coupling-assisted refrigeration in trapped ions}

\author{Chi-Chih Chen}
\thanks{These two authors contributed equally}
\email{l76011419@gs.ncku.edu.tw}
\affiliation{Institute of Atomic and Molecular Sciences, Academia Sinica, Taipei 10617, Taiwan}

\author{Yi-Cheng Wang}
\thanks{These two authors contributed equally}
\email{r09222006@ntu.edu.tw}
\affiliation{Department of Physics, National Taiwan University, Taipei 10617, Taiwan}
\affiliation{Institute of Atomic and Molecular Sciences, Academia Sinica, Taipei 10617, Taiwan}

\author{Chun-Che Wang}
\affiliation{Institute of Atomic and Molecular Sciences, Academia Sinica, Taipei 10617, Taiwan}

\author{H. H. Jen}
\email{sappyjen@gmail.com}
\affiliation{Institute of Atomic and Molecular Sciences, Academia Sinica, Taipei 10617, Taiwan}
\affiliation{Physics Division, National Center for Theoretical Sciences, Taipei 10617, Taiwan}

\date{\today}
\renewcommand{\r}{\mathbf{r}}
\newcommand{\f}{\mathbf{f}}
\renewcommand{\k}{\mathbf{k}}
\def\p{\mathbf{p}}
\def\q{\mathbf{q}}
\def\bea{\begin{eqnarray}}
\def\eea{\end{eqnarray}}
\def\ba{\begin{array}}
\def\ea{\end{array}}
\def\bdm{\begin{displaymath}}
\def\edm{\end{displaymath}}
\def\red{\color{red}}
\pacs{}
\begin{abstract}
Trapped ions can be cooled close to their motional ground state, which is imperative in implementing quantum computation and quantum simulation. Here we theoretically investigate the capability of light-mediated chiral couplings between ions, which enables a superior cooling scheme exceeding the single-ion limit of sideband cooling. Under asymmetric drivings, the target ion manifests the chiral-coupling-assisted refrigeration at the price of heating the others, where its steady-state phonon occupation outperforms the lower bound set by a single ion. We further explore the optimal operation conditions of the refrigeration, where a faster rate of cooling can still be sustained. Under an additional nonguided decay channel, a broader parameter regime emerges to support the superior cooling and carries over into the reciprocal coupling, suppressing the heating effect instead. Our results present a tunable resource of collective chiral couplings which can help surpass the bottleneck of cooling procedure and open up new possibilities in applications of trapped-ion-based quantum computer and simulator.  
\end{abstract}
\maketitle
\section{Introduction} 

Trapped-ion quantum computation \cite{Cirac1995} has reached a level of large-scale architecture \cite{Kielpinski2002, Pino2021, Shen2020}, where a high-performance universal quantum computer can be envisioned. In this scalable trapped-ion quantum computer, parallel zones of interactions and fast transport of ions can be integrated with high-fidelity gate operations \cite{Ballance2016, Gaebler2016} in multiple small quantum registers. One of the bottlenecks in achieving this feat is the cooling procedure \cite{Pino2021, Leibfried2003, Jordan2019} which aims to prepare the system in its motional ground state. Two commonly used cooling schemes in ions are sideband \cite{Diedrich1989, Monroe1995, Roos1999, Zhang2021} and electromagnetically-induced-transparency cooling \cite{Roos2000, Lechner2016, Feng2020, Qiao2021, Zhang2021_2, Wang2022}. Reaching the many-body ground state of ions is also essential in ensuring genuine quantum operations on these ionic registers, which can further enable simulations of other quantum many-body systems \cite{Buluta2009, Lanyon2011}. 

When multiple ions are involved in the cooling process, collective spin-phonon correlations arise owing to multiple scattering of light and recoil momentum \cite{Jordan2019, Shankar2019}, leading to effective dipole-dipole interactions between ions \cite{Cirac2000, Harlander2011}. This collective interaction \cite{Lehmberg1970} is ubiquitous in any light-matter interacting quantum interface \cite{Wang2022_mirror}, which can manifest a giant frictional force for atoms in an optical cavity \cite{Xu2016} or form optically bound pairs of atoms in free space \cite{Maximo2018, Gisbert2019}. The reciprocity nature of this light-induced dipole-dipole interactions can further be modified and controlled in an atom-waveguide interface \cite{Kien2005, Tudela2013, Kien2017, Solano2017, Chang2018, Corzo2019}, making the chiral quantum optical setup \cite{Gardiner1993, Carmichael1993, Stannigel2012, Luxmoore2013, Ramos2014, Arcari2014, Mitsch2014, Pichler2015, Sollner2015, Vermersch2016, Lodahl2017, Jen2019_selective, Jen2020_collective, Jen2021_crossover, Jen2020_subradiance, Jen2020_PRR} a novel scheme for exploration of motional refrigeration in optomechanical systems \cite{Xu2019, Lai2020}. 

\begin{figure}[b]
\centering
\includegraphics[width=0.48\textwidth]{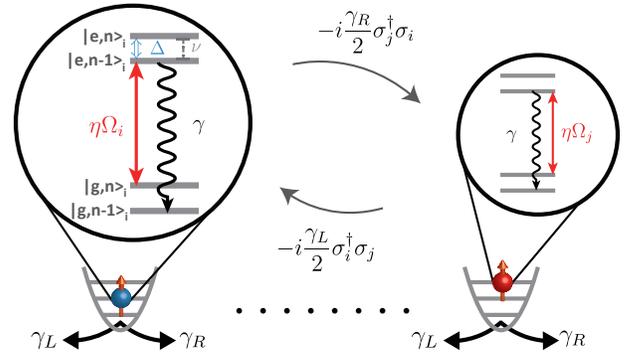}
\caption{A schematic plot of chiral couplings between ions. The ions are tightly confined in their respective trapping potentials under the sideband cooling scheme with an optimal cooling condition $\Delta$$=$$-\nu$, where $\Delta$ and $\nu$ are respectively the field detuning for the transition $|g,n\rangle$$\rightarrow$$|e,n\rangle$ and the trapping frequency. $\eta$ denotes the Lamb-Dicke parameter and $\Omega$ is Rabi frequency. An intrinsic decay rate for individual ion is $\gamma$, along with nonreciprocal decay channels $\gamma_L$ and $\gamma_R$ ($\gamma$$=$$\gamma_L$$+$$\gamma_R$). These left (L)- and right (R)-propagating decay rates represent the effective chiral couplings enabling spin-exchange hopping between $i$th and $j$th sites of ions.}\label{Fig1}
\end{figure}

Here we consider an ionic chain tightly confined in harmonic trapping potentials under the sideband cooling scheme and with collective chiral couplings, as shown in Fig. \ref{Fig1}. The chiral couplings between ions are employed to host spin-exchange hopping and nonreciprocal decay channels, where $\gamma_L\neq\gamma_R$. The effective coupling can be achieved either by moving the ions close to the waveguide \cite{Ong2020} where the guided modes mediate the long-range chiral couplings \cite{Vermersch2016} or by utilizing a chiral photonic quantum link in free space \cite{Grankin2018}. This setup leads to an unexplored territory of distinct heat exchange processes in cold ions. We note that it would be challenging to implement chiral couplings in ions through waveguide-mediated interactions owing to the uncontrollable surface charges on dielectrics. These charges lead to several adverse effects of unstable trapping or heating, which compromises optimized quantum operations \cite{Ong2020}. Nevertheless, ongoing efforts are in development to understand better the surface charge distribution and its stability, and these adverse effects can be mitigated if the waveguide can be discharged. 

In this article, we propose a novel cooling scheme that relaxes the assumption of single-particle spontaneous emission process. In essence, the intrinsic dissipation channel does not induce correlations between composite systems, and therefore many-atom cooling behavior can be attributed simply to single atom results. On the contrary, we introduce the resonant dipole-dipole interactions between atoms, which are universal in many light-matter interacting systems. Considering a one-dimensional atomic array subject to one-dimensional reservoir as in an atom-waveguide interface, we are able to further modify the dissipation process and its directionality, which allows tailored collective spin-exchange couplings and new parameter regimes for superior cooling performance. This results from the buildup and the dominance of spin-exchange process within the composite systems over the sideband cooling in a single ion, which enables a further heat removal. Furthermore, an extra nonguided channel we include can open a new paradigm to mitigate the heating effect at the reciprocal coupling, in essence to reduce the spin-phonon correlations which are otherwise more significant in heating. The tunable resource of collective chiral couplings we apply here can facilitate the motional ground state of ions and further push forward a large-scale and universal quantum computer employing trapped ions.

One of the crucial observations in our cooling scheme is the asymmetric driving condition. Under this condition, one of the ions in a one-dimensional atomic chain, the target ion, is driven with a relatively higher laser intensity, and the rest of them are the refrigerant ions acting as a reservoir of spin excitations and deexcitations for the target ion. With an additional asymmetry introduced in the nonreciprocal coupling strengths of $\gamma_R$ and $\gamma_L$, they further allow directional spin-exchange interactions, leading to an asymmetric heat transfer. This is the essence of refrigeration effect in multiple ions mediated with chiral couplings. As for the requirement of asymmetric driving condition, as long as we can sufficiently couple the refrigerant ions and the target ion by different intensities of laser fields, say only a fraction of one tenth or less for the refrigerant ones, we are safe in the superior cooling regime. Therefore, it does not matter how precise the coupling rates should be tuned as long as the asymmetric driving condition is satisfied. In our scheme, it would only require a relatively strong laser field on the target ion with weaker fields on the rest of the refrigerant ions in experiments to achieve our superior cooling performance. 

The paper is organized as follows. In Sec. II, we introduce the Hamiltonian of sideband cooling in composite ions with chiral couplings. In Sec. III, we present that light-mediated chiral couplings between ions enable a superior cooling scheme than the sideband cooling of a single ion. We find that the chiral-coupling-assisted refrigeration of the target ion can be feasible at a price of heating the other residual ones. In Sec. IV, we calculate the cooling dynamics and obtain the cooling rates. We investigate the effect of nonguided modes and multi-ion enhancement of cooling in Sec. V. In Sec. VI, we discuss the anomalous heating from ion traps and possible operations of our cooling scheme in quantum computation architecture. The Appendix presents the detail calculations of the steady-state phonon occupation in the target ion.     

\begin{figure*}[t]
\centering
  \begin{tabular}{cc}
    \includegraphics[width=0.98\textwidth]{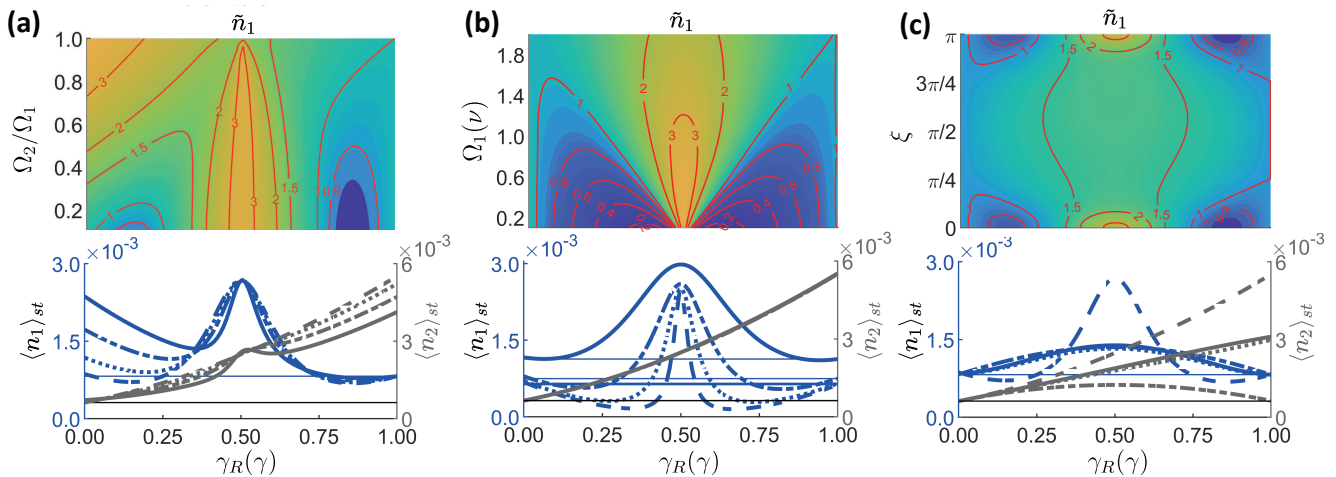}
  \end{tabular}
	\caption{Chiral-coupling-assisted refrigeration in the target ion. To identify the regimes of refrigeration or heating, we plot $\tilde n_i$ by comparing the result of respective single ions $\langle n\rangle^s_{\rm st}$ numerically. In all upper plots the cooler (warmer) colors represent lower (higher) $\tilde n_1$, and the lower panels show several horizontal cuts of the upper ones with exact values. We explore the effects of (a) $\Omega_2/\Omega_1$ with $\Omega_1=1\nu$, (b) $\Omega_1$ with $\Omega_2/\Omega_1=0.1$, and (c) $\xi$ with $\Omega_2/\Omega_1=0.1$, on the refrigeration of the target ion. Respective cuts are chosen at (a) $\Omega_2/\Omega_1=0.1$ (dashed), $0.3$ (dotted), $0.5$ (dash-dotted), and $0.7$ (solid), (b) $\Omega_1/\nu=0.2$ (dashed), $0.4$ (dotted), $0.8$ (dash-dotted), and $1.6$ (solid), and (c) $\xi=0$ (dashed), $\pi/4$ (dotted), $\pi/2$ (dash-dotted), $3\pi/4$ (solid) in blue lines. In all bottom plots, the corresponding $\langle n_2\rangle_{\rm st}$ in gray lines are shown for comparison, and they are almost overlapped in the case of the middle one. The horizontal lines are $\langle n\rangle^s_{\rm st}$ to guide the eye for the region that surpasses the single ion limit: the decay rate is chosen as $\gamma=0.1\nu$, and $\eta=0.04$.}\label{Fig2}
\end{figure*}

\section{Theoretical model} 

We consider a generic model of $N$ trapped ions with mass $m$ under standing wave sideband cooling \cite{Cirac1992} with chiral couplings in Lindblad forms \cite{Pichler2015}. The time evolutions of the density matrix $\rho$ of $N$ ions with quantized motional states $|n\rangle$ and an internal ground ($|g\rangle$) and excited states ($|e\rangle$) can be described by ($\hbar$$=$$1$) 
\bea
\frac{d \rho}{dt}=-i[H_{\rm LD}+H_L+H_R,\rho]+\mathcal{L}_L[\rho]+\mathcal{L}_R[\rho],\label{rho}
\eea
where $H_{\rm LD}$ for the sideband cooling in the Lamb-Dicke (LD) regime (in the first order of LD parameter $\eta$) reads
\bea
H_{\rm LD}&&=-\Delta\sum_{i=1}^N \sigma_i^\dag \sigma_i+\nu\sum_{i=1}^N a_i^\dag a_i\nonumber\\
&&+\frac{1}{2}\sum_{i=1}^N \eta\Omega_i(\sigma_i+\sigma_i^\dag)(a_i+a_i^\dag), \label{LD}
\eea
and the coherent and dissipative chiral couplings in the zeroth order of $\eta$ are, respectively, 
\bea
H_{L(R)} =&& -i\frac{\gamma_{L(R)}}{2} \sum_{\mu<(>)\nu}^N\left(e^{ik_s|r_\mu-r_\nu|} \sigma_\mu^\dag\sigma_\nu-\textrm{H.c.}\right)\label{HL}
\eea
and
\bea
\mathcal{L}_{L(R)}[\rho]=&&-\frac{\gamma_{L(R)}}{2} \sum_{\mu,\nu=1}^N e^{\mp ik_s(r_\mu-r_\nu)} \left(\sigma_\mu^\dag \sigma_\nu \rho + \rho \sigma_\mu^\dag\sigma_\nu \right.\nonumber\\
&&\left.-2\sigma_\nu \rho\sigma_\mu^\dag\right).\label{LR}
\eea
The laser Rabi frequency is $\Omega_i$ with a detuning $\Delta=\omega_L-\omega_{eg}$ denoting the difference between its central ($\omega_L$) and atomic transition frequencies ($\omega_{eg}$), and the dipole operators are $\sigma_\mu^\dag\equiv|e\rangle_\mu\langle g|$ with $\sigma_\mu=(\sigma_\mu^\dag)^\dag$. $\nu$ is the harmonic trap frequency with creation $a_i^\dag$ and annihilation operators $a_i$ in the Fock space of phonons $|n\rangle$, and LD parameter is $\eta=k_L/\sqrt{2m\nu}$ with $k_L\equiv\omega_L/c$. $k_s$ denotes the wave vector in the guided mode that mediates chiral couplings $\gamma_{L(R)}$, and we can use $\xi\equiv k_s |r_{\mu+1}-r_{\mu}|$ to quantify the light-induced dipole-dipole interactions associated with the relative positions of trap centers $r_\mu$ and $r_\nu$. 

The Lindblad forms in Eq. (\ref{rho}) take into account of spin-exchange processes between ions with nonreciprocal and long-range dipole-dipole interactions, and we use a normalized decay rate $\gamma=\gamma_R+\gamma_L$ to characterize the timescale of system dynamics. In the sideband cooling with $\eta\Omega,\gamma\ll\nu$ and the resolved sideband condition of $\Delta=-\nu$, the steady-state (st) phonon occupation in the case of a single ion can then be calculated as $\langle n\rangle_{st}^s$$\equiv$${\rm tr}(\rho_{st} a^\dag a)$$\propto$$(\gamma/\nu)^2$ with a cooling rate of $\mathcal{O}(\eta^2\Omega^2/\gamma)$ \cite{Cirac1992, Zhang2021} in the weak field regime. This presents that $\gamma$ determines the lower bound of phonon occupation, and a rate to reach this near motional ground state can be much smaller than $\gamma$. Next we explore the distinct cooling mechanism with the collective dipole-dipole interaction between every other ions in the sideband cooling scheme, where a superior cooling regime can be identified under an asymmetric driving condition $\Omega_i\neq\Omega_j$ on different ions. 

\section{Chiral-coupling-assisted cooling} 

We first demonstrate the chiral-coupling-assisted refrigeration in the case of two ions, which represents the essential element of interacting quantum registers. Whether there is refrigeration in these ions or not lies in their steady-state phonon occupations compared to their respective single-ion results without chiral couplings. We obtain the steady-state solutions by solving \(d\rho/dt=0\) in Eq. \ref{rho}, which is equivalent to finding a right eigen-matrix \(\rho_{st}\) with zero eigenvalue of Lindblad map, that is, \(\mathcal{L}[\rho_{st}]=0\) obtained from time-evolving solutions of \(\rho(t)=e^{t\mathcal{L}}[\rho(t=0)]\)~\cite{Carollo2021}. The steady-state solution of $\rho_{st}$ is also called the null space of Lindblad map, 
\bea
\rho_{st}= \text{Null}(\mathcal{L}), \label{NullRho}
\eea 
under the constraint of probability conservation $\text{Tr}(\rho_{st})=1$. The complete Hilbert space involves intrinsic spin and external motional degrees freedom, which we denote them as \(|\alpha, n\rangle_\mu \), where \(\alpha\in\{g,e\}\) denotes the ground and excited states for the $\mu$th ion and $n$ denotes the phonon number of phononic Fock states. Here we restrict \(n\in\{0,1\}\), which is valid when the dominant phononic Fock state is in the vicinity of the motional ground state. We note that for computing the null space of the Lindblad map, we convert the density matrix to Fock-Liouville space~\cite{Manzano2020}, which has a dimension equal to \(4^{2N}\) in our case, which leads to a computation complexity \(\mathcal{O}(4^{6N})\) by using singular-value-decomposition algorithms. This suggests a challenging task if not impossible in numerically simulating the case for $N=4$. 

In Fig. \ref{Fig2}, we numerically obtain the steady-state properties with up to a phonon number $n=1$, which is sufficient in the LD regime where $\langle n_i\rangle_{\rm st}\ll 1$. We use the normalized steady-state phonon occupation $\tilde n_i$$\equiv$\(\langle n_i\rangle_{\rm st}/\langle n\rangle^s_{\rm st}\) to present the cooling performance by comparing the result of respective single ions in a single-ion calculation versus $\gamma_R$, a right-propagating decay rate defined in Eq. (\ref{LR}) or schematically seen in Fig. \ref{Fig1}. The phonon occupation $\langle n\rangle^s_{\rm st}$ for a single ion has been calculated as $\langle n\rangle^s_{\rm st}$$\approx$$(\gamma/4\nu)^2$$+$$(\eta\Omega/\nu)^2/8$ under a weak or strong field regime \cite{Zhang2021}, and we also obtain them numerically in the bottom plots of Fig. \ref{Fig2} as a reference. The chiral-coupling-assisted cooling of the target ion (first ion) can be seen in the regions of $\tilde n_1<1$ in Fig. \ref{Fig2}(a) under an asymmetric driving. This is more evident when the driving field on the target ion is tuned weaker as shown in Fig. \ref{Fig2}(b). For a symmetric driving condition, refrigeration phenomenon never takes place. We also explore the effect of light-induced dipole-dipole interaction in Fig. \ref{Fig2}(c), where a superior cooling emerges at $\xi$ close to $\pi$ or $2\pi$. We find that the phonon occupation of the second ion is always larger than the one in a single-ion calculation, which acts as the refrigerant ion that always heats up while cools the target one. Under an asymmetrical driving condition, the refrigerant ion acts as a reservoir of spin excitations and deexcitations for the target ion. Therefore, the asymmetry between $\gamma_L$ and $\gamma_R$ further allows directional spin-exchange interactions, leading to an asymmetric heat transfer. We note that $\langle n_i\rangle_{\rm st}$ retrieves the single-ion results when $\gamma_R/\gamma=1$ and $0$ for the target and refrigerant ions, respectively. This results from the unidirectional coupling regime where spin-exchange couplings are forbidden, and thus spin-phonon correlations do not play a role in determining the steady-state properties.     

In Figs. \ref{Fig2}(a) and \ref{Fig2}(c), we find a moderate cooling performance of $\tilde n_1\lesssim 0.9$, which can further be pushed to below $0.2$ when $\Omega_1$ is made weaker in Fig. \ref{Fig2}(b). To understand the superior cooling parameter regimes in Fig. \ref{Fig2}(b), we trace over the phononic degrees of freedom of the refrigerant ion and investigate specifically the cooling performance in the target ion. Considering the perturbations of $\gamma^2$ and $\eta^2\Omega_1^2$ on an equal footing, we obtain the steady-state phonon occupation of the target ion by truncating to their first orders,
\bea
\langle n_1\rangle_{\rm st}&&\approx\frac{\gamma^2}{4\nu^2}\left(\frac{1}{2}-\frac{\gamma_R}{\gamma}\right)^2+\frac{\eta^2\Omega_1^2}{8\nu^2}\nonumber\\&&\times\left(\frac{\eta^2\Omega_1^2+2\gamma^2}{\eta^2\Omega_1^2+8\gamma^2(1/2-\gamma_R/\gamma)^2}\right),\label{n1}
\eea
which we calculate in detail in Appendix A. The excess heating for both the target and refrigerant ions shown in the below plots of Figs. \ref{Fig2}(a) and \ref{Fig2}(b) can be attributed to collective spin-exchange interactions especially under reciprocal couplings, contrary to the nonreciprocal couplings that can redirect the heating transfer between these two ions. This excess heating can as well be revealed in Eq. (\ref{n1}) for the target ion, where under the reciprocal coupling condition, the second bracket of Eq. (\ref{n1}) reaches its maximum and gives rise to the heating effect. The boundary that determines $\tilde n_1=1$ from Eq. (\ref{n1}) gives $\gamma_R=\gamma/2\pm\sqrt{3}\eta\Omega_1/2\sqrt{2}$, which delineates the onset of superior cooling and agrees well with numerical simulations in Fig. \ref{Fig2}(b). The linear dependence of $\gamma_R$ and $\Omega_1$ in the boundary indicates that excess cooling behavior happens symmetrically to the reciprocal coupling regime with a linear dependence of the driving field. This shows a competition between the laser driving field and the intrinsic spontaneous emission rate, where excess cooling emerges when $\eta\Omega_1\lesssim(2\gamma_R-\gamma)$. This also represents the dominance of spin-exchange process over the sideband cooling, which leads to a superior cooling performance. As for the symmetric dependence of $\gamma_R$ in the $\langle n_1\rangle_{\rm st}$ at small driving fields in the lower plot of Fig. \ref{Fig2}(b), this can be explained again by treating the refrigerant ion as a reservoir for spin-exchange interactions under the asymmetrical driving condition. The process of spin excitations and deexcitations of the target ion by spin-exchanging with the refrigerant ion effectively involves both the coupling strengths of $\gamma_R$ and $\gamma_L$, that is $\propto(\gamma_R/\gamma-1/2)(\gamma_L/\gamma-1/2)$, which leads to the symmetry in $\gamma_R$ or $\gamma_L$ with respect to $\gamma/2$. Under the condition of unidirectional coupling when $\gamma_R=\gamma$, $\langle n_1\rangle_{\rm st}$ again retrieves the single ion result $\langle n\rangle_{\rm st}^s$ as expected. 

\begin{figure}[t]
\centering
\includegraphics[width=0.48\textwidth]{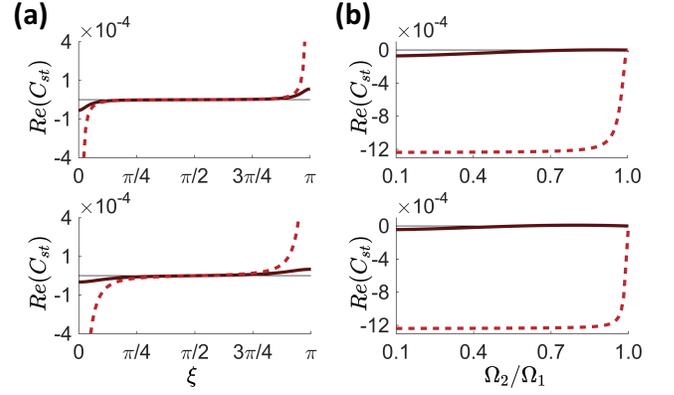}
\caption{Build-up of spin-spin correlation in chiral-coupling-assisted refrigeration. The nonclassical spin-spin correlations Re$(C_{st})$ are plotted as dependence of (a) $\xi$ at \(\Omega_2/\Omega_1=0.1\) and (b) $\Omega_2/\Omega_1$ at $\xi=2\pi$. In the upper and lower plots as comparisons, we choose $\Omega_1/\nu=0.2$ for $\gamma_R/\gamma=0.4$ (solid line) and $0.5$ (dashed line), and $\Omega_1/\nu=0.5$ for $\gamma_R/\gamma=0.25$ (solid line) and $0.5$ (dashed line), respectively. The solid-black line marks \({\rm Re}(\sigma_1^\dag \sigma_2)=0\). The decay rate of ions $\gamma$ is set to be the same as in Fig. \ref{Fig2}.}\label{Fig2_corr}
\end{figure}

We further identify three local extreme points in Eq. (\ref{n1}) as $\gamma_R/\gamma=0.5$ for one maximum $\langle n_1\rangle^{\rm max}_{\rm st}=\gamma^2/(4\nu^2)+\eta^2\Omega_1^2/(8\nu^2)$ which is always larger than $\langle n\rangle_{\rm st}^s$, and two equal minimums with corresponding values of $\gamma_R^{\rm min}$,
\bea
\langle n_1\rangle_{\rm st}^{\rm min}&&=\frac{\eta\Omega_1}{8\nu^2}\sqrt{\eta^2\Omega_1^2+2\gamma^2}-\frac{\eta^2\Omega_1^2}{32\nu^2},\\
\gamma_R^{\rm min}&&=\frac{\gamma}{2}\pm\frac{1}{2}\sqrt{\eta\Omega_1\sqrt{\eta^2\Omega_1^2+2\gamma^2}-\frac{\eta^2\Omega_1^2}{2}}.\label{n1min}
\eea
Interestingly, the local minimum $\langle n_1\rangle_{\rm st}^{\rm min}$ indicates a `mixing' effect of the driving field and the intrinsic decay rate, which results in $\tilde n_1^{\rm min}\approx 2\sqrt{2}\eta\Omega_1/\gamma$ when $\eta\Omega_1\rightarrow 0$. In this limit, the optimal condition of $\gamma_R^{\rm min}$ for this lower bound becomes close to $0.5\gamma$, which demonstrates the ultimate capability of reciprocal coupling in either cooling or heating, and strong spin-spin correlations therein. This can be illustrated in Fig. \ref{Fig2_corr}, where we show a build-up of finite spin-spin correlations $C_{st}=\langle\sigma_1^\dag\sigma_2\rangle-\langle\sigma_1^\dag\rangle\langle\sigma_2\rangle$ as a dependence of $\xi$ and asymmetric driving ratios. More significant correlations emerge in the heating regime, which we attribute to collective and reciprocal spin-exchange interactions. In the reciprocal coupling regime, the heat within the composite atomic system cannot be removed sufficiently, which can also be shown in the study of cooling rate in the next section. The reciprocal coupling regime leads to multiple reflections and transmissions of spin-exchange excitations and the resultant build-up of strong spin-spin correlations. This can be explained by the rising spin-phonon correlations introduced by sideband driving, which further induce stronger spin excitations translated into stronger spin-spin correlations via waveguide couplings. Meanwhile, the excess cooling regime at $\xi=\pi$ in Fig. \ref{Fig2_corr}(a) and at $\Omega_2/\Omega_1\approx 0.1$ in Fig. \ref{Fig2_corr}(b) shows a small but finite correlation. This suggests the essential role of finite spin-spin correlation which associates with collective spin-phonon coupling to remove extra heat, but not too much as in the heating regime. This also reflects an opened parameter window that allows excess cooling mechanism between the single atom or noninteracting regime with no correlations whatsoever and the heating regime with strong correlations. 

For a finite $\eta\Omega_1$, it gives room for a superior cooling performance than the single-ion case, which can be attributed to nonreciprocal spin-exchange couplings and distinct heat exchange processes. For typical parameters in Fig. \ref{Fig2} with $\Omega_1=0.1\nu$, the lower bound $\tilde n_1^{\rm min}\approx 0.11$, which shows an almost tenfold improvement than the single ion case, an order of magnitude advancement. We note that the lower bound that a single ion can achieve, however, suffers from an extremely slow cooling rate ($\propto$$\eta^2\Omega_1^2$). Next we show that the cooling rate of the target ion under chiral couplings, determined by a fitted overall timescale, can still surpass the single-ion case, but a longer time is needed for reaching the steady state owing to a small $\eta\Omega_1$.    

\section{Cooling rate} 

\begin{figure}[b]
\centering
\includegraphics[width=0.48\textwidth]{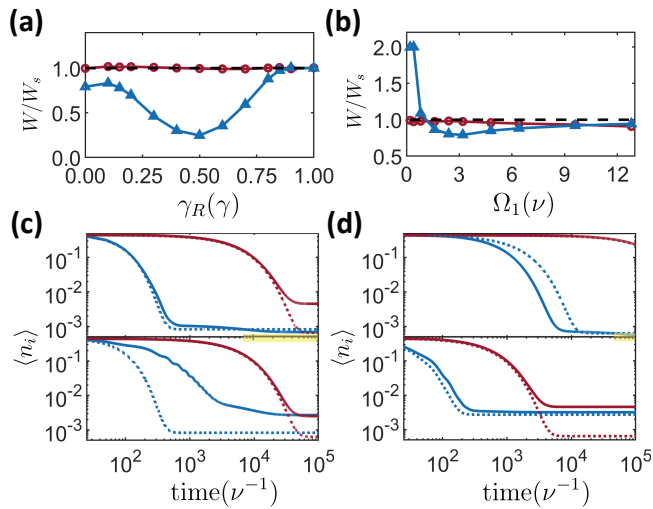}
\caption{Cooling rates $W$ of the target and refrigerant ions. The condition for the initial thermal ensemble of ions is taken as $n_0= 0.7$ and a truncation of phonon number to $n=4$. All cooling rates of the target (blue-$\blacktriangle$) and refrigerant ions (red-$\bullet$) are compared to their respective single-ion results $W_s$ (dashed lines) as a dependence of (a) $\gamma_R$ with $\Omega_1/\nu=1$ and (b) $\Omega_1$ with $\gamma_R/\gamma=0.85$, where both plots take $\Omega_2/\Omega_1=0.1$ and $\xi=2\pi$. The corresponding time evolutions of phonon occupations (blue- and red-solid lines) in (a) and (b) are shown in (c) and (d), respectively, for $\gamma_R/\gamma=0.85$, $0.5$, and $\Omega_1/\nu=0.2$, $3.2$, in the upper and lower plots. The respective single-ion results (dashed lines) are plotted for comparisons. The refrigeration effect initiates before and after the time $\sim 10^4$ ($\nu^{-1}$) in (c) and (d) with yellow-shaded areas. The $\gamma$ is set to be the same as in Fig. \ref{Fig2}, and the inset plots in (c) and (d) are normalized $\tilde n_1$ for an identification of the time crossing $\tilde n_1=1$ when cooling initiates and sustains.}\label{Fig3}
\end{figure}

In numerically simulating the time dynamics of the phonon occupations for both ions as shown in Fig. \ref{Fig3}, we assume the initial state of the trapped ions in a thermal state \cite{Zhang2021, Roos2000_2}, 
\bea
\rho(t=0)=\Pi_{\mu=1}^N \sum_{n=0}^\infty \frac{n_0^n}{(n_0+1)^{n+1}} |g, n\rangle_\mu\langle g, n|,
\eea
where $n_0$ is an average phonon number for both ions. We use $n_0\lesssim 1$ with a finite truncation of the motional states to guarantee the convergence in numerical simulations. To quantify the cooling behaviors, we use an exponential fit for the timescale to reach $\langle n_i\rangle_{\rm st}$ with a function of $ae^{-bt}+\langle n_i\rangle_{\rm st}$ for arbitrary constants $a$ and $b$. We then obtain the corresponding cooling rates $W=b$, which generally gives an overall timescale of the cooling process. 

In Figs. \ref{Fig3}(a) and \ref{Fig3}(b), we show the fitted cooling rates comparing the respective single-ion results and corresponding time evolutions in Figs. \ref{Fig3}(c) and \ref{Fig3}(d). The different panels of Figs. \ref{Fig3}(c) and \ref{Fig3}(d) correspond to the time evolutions with the parameter regimes in Figs. \ref{Fig3}(a) and \ref{Fig3}(b), respectively, where we have chosen the cooling and heating cases of the target ion in the upper and lower panels as comparisons. For the refrigerant ion, the cooling rate does not change significantly and behaves similarly to the single ion case with a rate $\propto$$\eta^2\Omega_2^2$, showing a rather prolonged time dynamics owing to an asymmetric setting of the driving fields. Meanwhile, a faster cooling rate emerges for the target ion when $\gamma_R\approx 0.85$ and $\Omega_1/\nu\lesssim 1.5$, as shown in Fig. \ref{Fig3}(b). The time region when the target ion surpasses the single-ion limit can be seen in Figs. \ref{Fig3}(c) and \ref{Fig3}(d), where the refrigeration effect shows up at a later stage than the single ion case. The time for establishing refrigeration appears approximately ten times longer than the one a single ion reaches its steady state, which suggests the price one has to pay in applying this superior cooling scheme under chiral couplings. 

The slow rates of $W$ in Fig. \ref{Fig3}(a) at $\gamma_R/\gamma\sim 0.5$ reflects a delay from multiple exchanges of spin excitations and phonon occupations, while a retrieved rate of single ion emerges again in the unidirectional coupling regime. As $\Omega_1$ increases in Fig. \ref{Fig3}(b), both cooling rates approach respective single-ion cases, which depend on $\gamma/[2(1+n_0)]$ bounded by $\gamma$ \cite{Zhang2021}. The slow cooling rates at the reciprocal coupling regime can be attributed to a lack of directionality in dissipation. This leads to a slow spread of spin diffusion \cite{Jen2021_bound, Jen2022_correlation} and associated stagnant removal of phonon, in addition to the buildup of spin-spin correlations owing to the collective nature of nonreciprocal couplings between these constituent atoms. We note as well that the reciprocal coupling regime allows a more significant interference in spin populations, which is highly related to the multiple reflections and transmissions in spin exchanges before they relax as time evolves. This could be one of the reasons why the system takes a longer time to reach the steady state in Fig. \ref{Fig3}(a).    

\section{Effect of nonguided decay and multi-ion case} 

Here we introduce an additional nonguided mode on top of the guided nonreciprocal couplings. This makes our system away from a strong coupling regime but closer to a realistic setting, where unwanted decays can be unavoidable \cite{Lodahl2017}. The nonguided decay rate $\gamma_{ng}$ can simply be cast into Eq. (\ref{rho}) in a form of 
\bea
\mathcal{L}_{ng}[\rho]=-\frac{\gamma_{ng}}{2}\sum_{\mu=1}^N\left(\sigma_\mu^\dag \sigma_\mu \rho + \rho \sigma_\mu^\dag\sigma_\mu -2\sigma_\mu \rho\sigma_\mu^\dag\right).
\eea  
A parameter of $\beta\equiv\gamma/(\gamma+\gamma_{ng})$ can quantify the crossover from a strong coupling ($\beta=1$) to a purely noninteracting regime ($\beta=0$). 

\begin{figure}[t]
\centering
\includegraphics[width=0.48\textwidth]{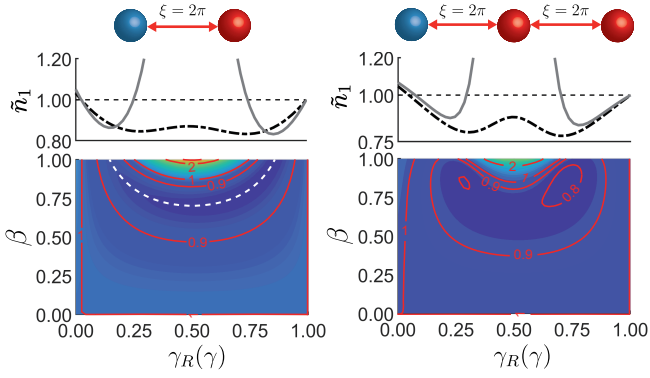}
\caption{Nonguided mode in cooling the target ion. The nonguided decay rate $\gamma_{ng}$ is introduced in the cases of two and three ions with an equal interparticle distance at $\xi=2\pi$, where \(\beta\equiv\gamma/(\gamma+\gamma_{ng}) \) indicates the portion of decay to the guided mode. Similar shading color is used in respective lower panels as in Fig. \ref{Fig2} with the parameters of $\Omega_2/\Omega_1=0.1$, $\Omega_1=1\nu$, and $\gamma=0.1\nu$. The upper panels present some cuts in the lower ones at \(\beta=1\) (solid), \(\beta=0.8\) (dash-dotted), and \(\beta=0\) (dashed). A dashed line in the lower plot of the two-ion case represents a local minimum predicted from an analytical derivation in Appendix A.}\label{Fig4}
\end{figure}

As shown in Fig. \ref{Fig4}, we find a broader parameter regime of $\beta$ that can sustain the better cooling performance where $\tilde n_1<1$ and further reduce its local minimum of phonon occupations. More surprisingly, the heating behavior at the reciprocal coupling of $\gamma_R/\gamma=0.5$ can be suppressed and turned to cooling instead with $\beta\lesssim 0.9$. This is manifested as well in the case of three ions under asymmetric drivings, where the target ion can still present a superior cooling behavior with an even lower $\tilde n_1^{\rm min}$ using two refrigerant ions. The crescent-like region of low $\tilde n_1$ in the case of two ions can be analyzed by tracing over the refrigerant ion's motional states. An analytical prediction of the local minimums, which results from a quartic equation of $\beta^2(\gamma_R/\gamma)^2$ in Appendix A.1, is shown on top with this region. This leads to two local minimums for a fixed and finite $\beta$ and a continuation of $\tilde n_1^{\rm min}$ at $\beta=1$ toward the parameter regimes of $\beta<1$ and $\gamma_R=0.5\gamma$, which provides a route to superior cooling even under a finite $\gamma_{ng}$. The reason why the superior cooling can be allowed here might be due to the extra dissipative channel that mitigates the effect of reciprocal couplings. This extra dimension of nonguided mode provides the possibility for the composite system to explore between the regimes with highly correlated spin-phonon couplings at $\gamma_R=\gamma_L$ with $\beta=1$ and purely noninteracting ones at $\beta=0$. Since the cooling performance of the target ion reduces to the single-ion result at $\beta=0$, naturally and as expected a superior cooling regime would emerge in between for a finite $\beta$. We can also attribute these new parameter regions for cooling to a reduction of spin-spin correlations, which are otherwise more evident in the heating regime as shown in Figs. \ref{Fig2} and \ref{Fig2_corr}. Essentially, the role of the nonguided mode here makes the composite system less susceptible to the collective spin-exchange interactions which are augmented the most at the reciprocal coupling regime.   

For the case of multiple ions under asymmetric drivings, we are able to take the partial trace of the motional degrees of freedom in the refrigerant ions by assuming the laser driving strengths on them are small enough. This leads to a reduced Hilbert space spanned by complete internal and
motional states of the target ion and only the internal states of refrigerant ones. Although the relative location of the target ion to other refrigerant ones can matter as seen from Eqs. (\ref{HL}) and (\ref{LR}) under chiral couplings, we have checked that the configurations of the target ion in an ionic periodic array of $N=3$ is irrelevant under the asymmetric driving condition, that is, $\bigl\langle n_{1}\bigr\rangle_{\text{st}}$ is the same for the target ion in the end or the middle of the chain when the interparticle separation is chosen as $\xi=2\pi$. Therefore, we consider that the target ion locates at the leftmost site of an $N$-ion chain without loss of generality. 

\begin{figure*}[t]
\centering{}\includegraphics[width=\textwidth]{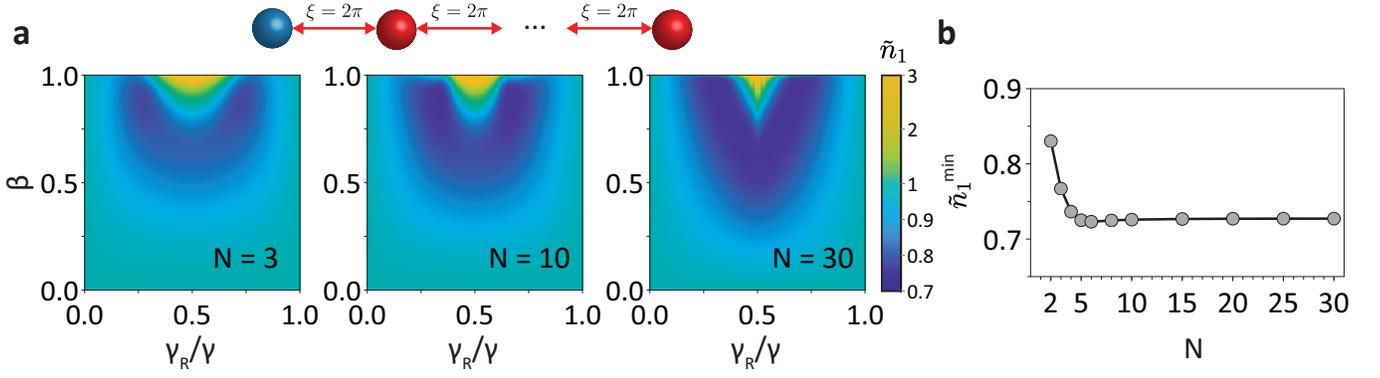}
\caption{\label{fig:N-ion}Steady-state phonon occupation number of the target
ion as a function of $\beta$ and $\gamma_{R}/\gamma$ at multi-ion
case. (a)~ Numerically calculated $ \tilde{n}_{1}$ under the asymmetric driving and the interparticle distances chosen as multiples of 2$\pi$. (b)~ Numerically calculated global minimum of $\tilde{n}_{1}^{\text{min}}$. The parameters used here are \(\eta=0.04\), \(\Omega=1\nu\), and \(\Gamma=0.1\nu\).}
\end{figure*}

We proceed by keeping the density matrix elements whose leading terms are up to the order of $\gamma^2/\nu^2$ and $\eta^2\Omega^2/\nu^2$. We find that they can be selected by the following two rules. One is the Hamming distance between the specific density matrix element and that of many-body ground state (e.g., $\rho_{g0ggg0gg}$ for three-ion case) is not greater than two. The other is that the row and column indices of the density matrix elements can only contain at most one excited state, where $\bigl|e\bigr\rangle$ and $\bigl|n=1\bigr\rangle$ are treated as excited states. However, there is an exception for $\rho_{e1g...g,e1g...g}$ which should be included since it represents the population in $\bigl|e1\bigr\rangle_{1}$ state, which is $\mathcal{O}(\eta^{2}\Omega^{2})$ due to the driving on the target ion. With these conditions, we find that the following relationships still hold as in Eq. (\ref{relation}), 
\begin{eqnarray}
\rho_{e1g...g,e1g...g} & = & \frac{\eta^{2}\Omega^{2}}{16\nu^{2}}\rho_{g0g...g,g0g...g},\\
\rho_{\underbrace{{\scriptstyle g1g...geg...g}}_{(i+2)\text{th index is }e},g0g...g} & = & -\frac{i\gamma_{R}\eta\Omega}{8\nu^{2}}\rho_{g0g...g,g0g...g},\\
\rho_{e1g...g,g0g...g} & = & -\frac{4\nu-i\Gamma}{16\nu^{2}}\eta\Omega\rho_{g0g...g,g0g...g},
\end{eqnarray}
where the first two indices in the row and column ones represent the internal and motional state of the target ion, and the $(i+2)$th index stands for the internal state of the $i$th refrigerant ion for $i\in[1,N-1]$.

Next, we construct the multi-ion generalization of Eq. (\ref{a3}). Here we categorize these undetermined density matrix elements according to the indices of the target ion as follows,
\begin{eqnarray}
B_{i} & = & \rho_{g1g...g,\underbrace{{\scriptstyle g0g...geg...g}}_{(i+2)\text{th index is }e}}=-\rho_{\underbrace{{\scriptstyle g0g...geg...g}}_{(i+2)\text{th index is }e},g1g...g},\\
C_{i} & = & \rho_{e0g...g,\underbrace{{\scriptstyle g0g...geg...g}}_{(i+2)\text{th index is }e}}=\rho_{\underbrace{{\scriptstyle g0g...geg...g}}_{(i+2)\text{th index is }e},e0g...g},\\
D_{ij} & = & \rho_{\underbrace{{\scriptstyle g0g...geg...g}}_{(i+2)\text{th index is }e},\underbrace{{\scriptstyle g0g...geg...g}}_{(j+2)\text{th index is }e}}=\rho_{\underbrace{{\scriptstyle g0g...geg...g}}_{(j+2)\text{th index is }e},\underbrace{{\scriptstyle g0g...geg...g}}_{(i+2)\text{th index is }e}},\nonumber\\
\end{eqnarray}
where $D_{ji}=D_{ij}$. The above represent spin-phonon and spin-spin correlations between refrigerant and the target ions, and spin-spin correlations within refrigerant ones, respectively. Combining the above variables with other undetermined variables, such as $A=\rho_{g1g...g,e0g...g}=-\rho_{e0g...g,g1g...g}$, $\rho_{e0g...g,e0g...g}$, and $\rho_{g1g...g,g1g...g}$, we obtain the following coupled equations
\begin{eqnarray}
0  = && -2i\eta\Omega A-2\Gamma\rho_{e1g...g,e1g...g},\label{eq:3E-1}\\
0  = && \Gamma B_{i}+2\gamma_{R}A+2\gamma_{R}\sum_{j=1}^{i-1}B_{j}\nonumber\\
&&+2\gamma_{L}\sum_{j=i+1}^{N-1}B_{j}+i\eta\Omega C_{i},\\
0  = && 2\Gamma C_{i}+2\gamma_{R}\sum_{j=1}^{i-1}C_{j}+2\gamma_{L}\sum_{j=i+1}^{N-1}C_{j}\nonumber\\
&&+2\gamma_{L}\sum_{j=1}^{N-1}D_{ji}+i\eta\Omega B_{i}+2\gamma_{R}\rho_{e0g...g,e0g...g},\\
0  = && \Gamma D_{ij}+\gamma_{R}\sum_{k=1}^{i-1}D_{kj}+\gamma_{L}\sum_{k=i+1}^{N-1}D_{kj}\nonumber\\
&&+\gamma_{R}\sum_{k=1}^{j-1}D_{ik}+\gamma_{L}\sum_{k=j+1}^{N-1}D_{ik}+\gamma_{R}(C_{i}+C_{j}),\\
0  = && 2\Gamma\rho_{e0g...g,e0g...g}+4\gamma_{L}\sum_{j=1}^{N-1}C_{j}+2i\eta\Omega A,\label{eq:3E-2}
\end{eqnarray}
where we have $N(N+3)/2$ variables, i.e., $A$, $B_{i}$, $C_{i}$, $D_{ij}$~($i\leq j$, real symmetric matrix), and $\rho_{e0g...g,e0g...g}$. They can be solved numerically in terms of $\rho_{e1g...g,e1g...g}$ or equivalently $\rho_{g0g...g,g0g...g}$, and finally we obtain $\rho_{g1g...g,g1g...g}$ from
\begin{eqnarray}
0 = && i\eta\Omega(\rho_{e0g...g,e0g...g}-\rho_{g1g...g,g1g...g})+\Gamma A\nonumber\\
&&+2\gamma_{L}\sum_{j=1}^{N-1}B_{j}.
\end{eqnarray}

We note of a tremendous reduction of the number of coupled equations in Eqs. (\ref{eq:3E-1}-\ref{eq:3E-2}), which gives a power law $\mathcal{O}(N^{2})$ complexity compared to the exponential one $\mathcal{O}(4^{2N})$ in full Hilbert space. This allows us to calculate chiral-coupling-assisted cooling in the ionic chain with dozens of ions. In Fig. \ref{fig:N-ion}(a), we show three representative demonstrations of $\tilde{n}_{1}$ at $N=3$, $10$, and $30$, where the region within $\tilde{n}_{1}\lesssim 0.8$ becomes wider as $N$ increases. The $N$ dependence of the global minimum of $\tilde{n}_{1}^{\text{min}}$ is shown in Fig. \ref{fig:N-ion}(b), which saturates to a lower bound at $\tilde{n}_{1}\approx0.725$ after $N\geq5$. This presents the potentiality in multi-ion-assisted cooling via collective chiral couplings. 

\section{Discussion and Conclusion}

We have shown theoretically that the chiral couplings introduced in the trapped-ion system enable a better cooling performance than a single ion in the sideband cooling. This light-mediated chiral coupling between ions manifests a resource with capability to achieve a superior cooling scheme that surpasses the lower bound of the steady-state phonon occupation a single ion can allow. The chiral-coupling-assisted refrigeration in two and three ions can be useful in a large-scale quantum computer composed of multiple small entities of ions without compromising the cooling rates. When $\gamma/2\pi=20$ MHz is used in our results, it gives a cooling time of $10^5 (\nu^{-1})$ within $100~\mu$s, which is feasible in several typical platforms of $^9$Be$^+$ \cite{Shankar2019}, $^{40}$Ca$^+$ \cite{Staanum2004}, $^{172}$Yb$^+$ \cite{Kielpinski2006}, or $^{171}$Yb$^+$ ions \cite{Feng2020}. In conclusion, our results present a distinctive control over the motional ground states with tunable chiral couplings and provide new insights in getting around the cooling barrier in trapped-ion-based applications of quantum computer and simulator. Last but no least, the scheme we consider here can also be implemented with optical tweezers in a scalable ion crystal for high-performance gate operations \cite{Olsacher2020, Mazzanti2021}.

We note that an anomalous heating is unavoidable in ion traps owing to the electric field noise from the electrode surfaces. The anomalous heating could be an issue in our new cooling scheme when it becomes the dominating factor. This, however, can be lessened by lowering the electrode temperature \cite{Deslauriers2006}, applying surface plasma cleaning \cite{McConnell2015}, or increasing the axial trap frequency with higher trapping heights \cite{Deslauriers2006, Boldin2018}. Considering $\gamma/2\pi=20$ MHz for the decay rate again, we estimate that a $10^{-3}$ phonon number gives a temperature $T\approx 1.3\times 10^{-3}$ Kelvin (K) \cite{Leibfried2003}. Within a cooling time of $100~\mu$s, we can further estimate the comparable anomalous heating rate as $T/(100~\mu\rm{s})\approx 13$ K$/$s, which sets the lower bound that would compromise our cooling scheme. Again, the anomalous heating rate can be made much smaller than the estimated bound $13$ K$/$s by tuning the axial frequency and ion-surface separation, which can be as low as $0.01$ K$/$s and within experimental reach \cite{Boldin2018}. 

Finally, for quantum computation protocol using our cooling scheme with multiple ions, we resort to the trapped-ion quantum charge-coupled device as quantum computer architecture \cite{Pino2021}. In the similar spirits of using parallel interaction zones, our multi-ion cooling scheme can be implemented in parallel as well, which would prepare the target ions close to the motional ground state even in the case of two ions. This coincides with the design using a small-ion crystal, which presents a better performance in state preparation or gate operation owing to its high controllability. We then can collect all the target ions into the interaction zone after cooling procedure via an adiabatic ion transport. Presumably within a small-ion crystal, we can save some error and time budget in quantum computation from our proposed scheme. For more ions, as shown in Fig. \ref{fig:N-ion}(b), the surpassing cooling performance saturates as $N$ increases, and these many ions would experience unexpected heating owing to system complexities of electric field noises or laser field fluctuations. As for design spirits of small-ion quantum registers, our multi-ion enhancement in cooling could be compromised, but it is still good to know that already a reasonable superior cooling performance can be achieved with less than three or four ions in our scheme. Essentially, our cooling scheme offers an alternative method to go around the cooling protocol bottleneck, which helps improve the quantum computation architecture.

\section*{ACKNOWLEDGMENTS}
We acknowledge support from the Ministry of Science and Technology (MOST), Taiwan, under the Grant No. MOST-109-2112-M-001-035-MY3. We are also grateful for support from TG 1.2 and TG 3.2 of NCTS and inspiring discussions with G.-D. Lin. 

\appendix
\section{Analytical form of the steady-state occupation of the target ion}

In chiral-coupling-assisted cooling of two ions, the Hilbert space dimension is $16$ with $256$ coupled linear equations, which hardly gives insightful results analytically. To explore the optimal condition for the target ion in the steady state, we perform partial trace to the refrigerant ion with respect to the motional degree of freedom ($a_{2}$), which diminishes the dimension of the Hilbert space to $8$. This is valid if the laser driving strength of the refrigerant ion is much smaller than that of the target ion. In this Appendix, we replace \(\Omega_1\) by \(\Omega\) for simplicity, and we define the total decay rate as \(\Gamma=\gamma_R+\gamma_L+\gamma_{ng}\), which is fixed by the intrinsic decay rate of ion $\Gamma$.

	\begin{figure*}[t]
		\centering
  \begin{tabular}{cc}
		\includegraphics[width=172mm]{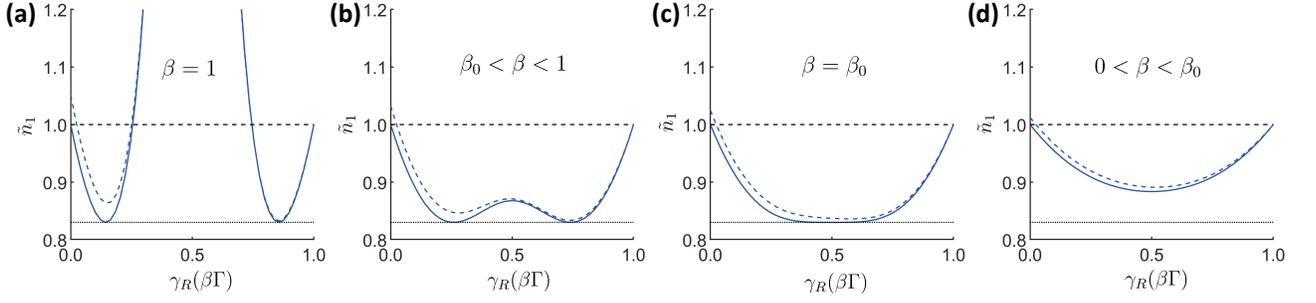}
		\end{tabular}
		\caption{Normalized steady-state phonon occupation of target ion. The blue solid lines and blue dashed lines display the results from Eq.~\eqref{eq:nstAnalytical} and the numerical simulation, respectively. From left to right panels, the corresponding total coupling efficiencies \(\beta\) are (a)~1, (b)~0.8, (c)~0.7, and (d)~0.5. The other parameters are \(\eta=0.04\), \(\Omega=1\nu\), \(\Gamma=0.1\nu\), which lead to \(\beta_0 \approx 0.7\). The Rabi frequency of the laser drive to the refrigerant ion for the blue dashed line is \(0.1\nu\). The horizontal dashed and dotted lines show the \(\langle n_1 \rangle_{st}^{s}\) and \(\langle n_1 \rangle_{st}^{min}.\) }
		\label{fig:Fig3}
	\end{figure*}

Now the dynamics of this system can be determined by the reduced density matrix $\text{Tr}_{a_{2}}(\rho)$. Since we focus on solving \(\langle n_1 \rangle_{st}\), the number of equations required can be further reduced to $20$. These equations generally involve the steady-state density matrix elements of $\rho_{\mu_{1}n_{1}\mu_{2}\nu_{1}m_{1}\nu_{2}}=\langle\mu_{1},n_{1};\mu_{2}|\text{Tr}_{a_{2}}\rho|\nu_{1},m_{1};\nu_{2}\rangle$. In the resolved sideband cooling under Lamb-Dicke regime $\Delta=-\nu$ along with the condition \(e^{ikd}=1\), we can take advantage of the fact that $\rho_{g0gg0g}$ is $\mathcal{O}(1)$, and $\gamma^{2}/\nu^{2}$ and $\eta^{2}\Omega^{2}/\nu^{2}$ are much smaller than one. As a result, we neglect those density matrix elements whose leading term is higher than second order, such as \(\rho_{e1ge0e}\), \(\rho_{g0ge0e}\), \(\rho_{g1ee0e}\), \(\rho_{e1ee0g}\), \(\rho_{e0ge1e}\), \(\rho_{g0ee1e}\), \(\rho_{g1ge1e}\), \(\rho_{e0ee1g}\), \(\rho_{g1ee1g}\), \(\rho_{e1eg0e}\), \(\rho_{e0eg0g}\), \(\rho_{e0eg1e}\), \(\rho_{e1gg1e}\), \(\rho_{e1eg1g}\), \(\rho_{e0ee0e}\), \(\rho_{e1ee1e}\), and \(\rho_{g1eg1e}\). This leads to	
	\begin{equation}
	\begin{split}
		0 = & \eta\Omega(\rho_{e0gg1g}-\rho_{g1ge0g})+2i\gamma\rho_{e0ge0g}\\
		&+2i\gamma_{L}(\rho_{g0ee0g}+\rho_{e0gg0e}),\\
		0 = & -\eta\Omega\rho_{g0eg1g}-2i\gamma_{L}\rho_{g0eg0e}-2i\gamma_{R}\rho_{e0ge0g}\\
		&-2i\gamma\rho_{g0ee0g},\\
		0 = & \eta\Omega(\rho_{e0ge0g}-\rho_{g1gg1g})-2i\gamma_{L}\rho_{g1gg0e}\\
		&-i\gamma\rho_{g1ge0g},\\
		0 = & -2i\gamma_{R}(\rho_{e0gg0e}+\rho_{g0ee0g})-2i\gamma\rho_{g0eg0e},\\
		0 = & \eta\Omega\rho_{e0gg0e}-2i\gamma_{R}\rho_{g1ge0g}-i\gamma\rho_{g1gg0e},\\
		0 = & \eta\Omega(\rho_{e0gg1g}-\rho_{g1ge0g})+i\gamma\frac{\eta^{2}\Omega^{2}}{8\nu^{2}}\rho_{g0gg0g},
		\end{split}
	\end{equation}
where we have used the following relationships, 
	\begin{align}\label{relation}
	\begin{split}
		\rho_{e1ge1g}  =&  \frac{\eta^{2}\Omega^{2}}{16\nu^{2}}\rho_{g0gg0g},\\
		\rho_{g1eg0g}   =&  -\frac{i\gamma_{R}\eta\Omega}{8\nu^{2}}\rho_{g0gg0g},\\
		\rho_{e1gg0g}   =&  -\frac{4\nu-i\Gamma}{16\nu^{2}}\eta\Omega\rho_{g0gg0g}.
		\end{split}
	\end{align}
	Finally we have the following density matrix elements expressed in terms of \(\rho_{g0gg0g}\), 
	\begin{align}\label{a3}
	\begin{split}
	\rho_{e0gg1g}=&-\rho_{g1ge0g}=-i\Gamma\frac{\eta\Omega}{16\nu^{2}}\rho_{g0gg0g},\\
	\rho_{e0gg0e}=&\rho_{g0ee0g}=-\frac{\Gamma}{2\gamma_{R}}\rho_{g0eg0e},\\
		\rho_{g0eg0e}  = & \frac{\gamma_{R}^{2}}{\frac{\Gamma^{2}}{4}-\gamma_{R}\gamma_{L}+\frac{\eta^{2}\Omega^{2}}{8}}\frac{\eta^{2}\Omega^{2}}{16\nu^{2}}\rho_{g0gg0g},\\
		\rho_{e0gg0e}  = & -\frac{\Gamma\gamma_{R}}{\frac{\Gamma^{2}}{4}-\gamma_{R}\gamma_{L}+\frac{\eta^{2}\Omega^{2}}{8}}\frac{\eta^{2}\Omega^{2}}{32\nu^{2}}\rho_{g0gg0g},\\
		\rho_{e0ge0g}  = & \frac{\frac{\Gamma^{2}}{4}+\frac{\eta^{2}\Omega^{2}}{8}}{\frac{\Gamma^{2}}{4}-\gamma_{R}\gamma_{L}+\frac{\eta^{2}\Omega^{2}}{8}}\frac{\eta^{2}\Omega^{2}}{16\nu^{2}}\rho_{g0gg0g},\\
		\rho_{g1gg0e}  = & i\frac{\frac{\eta^{2}\Omega^{2}}{8}-\frac{\Gamma^{2}}{4}+\gamma_{R}\gamma_{L}}{\frac{\Gamma^{2}}{4}-\gamma_{R}\gamma_{L}+\frac{\eta^{2}\Omega^{2}}{8}}\frac{\gamma_{R}\eta\Omega}{8\nu^{2}}\rho_{g0gg0g},\\
		\rho_{g1ge0g}  = & i\Gamma\frac{\eta\Omega}{16\nu^{2}}\rho_{g0gg0g},\\
		\rho_{g1gg1g}  = & \frac{1}{16\nu^{2}}\rho_{g0gg0g}\Biggl[(\Gamma^{2}-4\gamma_{R}\gamma_{L})\\
		&+\eta^{2}\Omega^{2}\frac{\frac{\Gamma^{2}}{4}+\gamma_{R}\gamma_{L}+\frac{\eta^{2}\Omega^{2}}{8}}{\frac{\Gamma^{2}}{4}-\gamma_{R}\gamma_{L}+\frac{\eta^{2}\Omega^{2}}{8}}\Biggr].
		\end{split}
	\end{align}
	
The steady-state occupation for the target ion can therefore be derived as (\(\rho_{g0gg0g}\approx 1\))
	\begin{eqnarray}
	\bigl\langle n_{1}\bigr\rangle_{\text{st}} = && \rho_{e1ee1e}+\rho_{e1ge1g}+\rho_{g1eg1e}+\rho_{g1gg1g},\nonumber \\
	 = && \frac{\Gamma^{2}}{16\nu^{2}}+\frac{\eta^{2}\Omega^{2}}{8\nu^{2}}-\frac{\gamma_{R}\gamma_{L}}{4\nu^{2}}\nonumber\\
	&&+\frac{\eta^{2}\Omega^{2}}{\eta^{2}\Omega^{2}+2\Gamma^{2}-8\gamma_{R}\gamma_{L}}\frac{\gamma_{R}\gamma_{L}}{\nu^{2}},\label{eq:nstAnalytical}
	\end{eqnarray}	
	 where the first two terms are the steady-state phonon occupation of a single ion cooling, and the remaining terms are the modifications arised from the chiral couplings. The comparison between the prediction from Eq.~\eqref{eq:nstAnalytical} and the numerical simulation is shown in Fig.~\ref{fig:Fig3}. The blue dashed lines represent the numerical results without partial tracing out the refrigerant ion's motional degree of freedom, and the blue solid lines show our analytical results. The blue solid lines display a mild deviation from the numerical result on the side \(\gamma_R<0.5\) since the simulation results include the influence of finite laser driving of the refrigerant ion, which causes the asymmetry of the \(\langle n_1 \rangle_{\rm st}\)--\(\gamma_R\) curve.
	 
\subsection{Minimal phonon occupation of the target ion}
	
From Eq.~\eqref{eq:nstAnalytical}, the minimal phonon occupation of target ion can be obtained as
	 \begin{align}
	 	\bigl\langle n_{1}\bigr\rangle_{\text{st}}^{\text{min}} & =\bigl\langle n_{1}\bigr\rangle_{\text{st}}^{s}-\frac{1}{32\nu^{2}}\Bigl(\sqrt{\eta^{2}\Omega^{2}+2\Gamma^{2}}-2\eta\Omega\Bigr)^{2},\nonumber \\
	 	& =\frac{\eta\Omega}{8\nu^{2}}\sqrt{\eta^{2}\Omega^{2}+2\Gamma^{2}}-\frac{\eta^{2}\Omega^{2}}{32\nu^{2}},\label{eq:100}
	 \end{align}
	where the minimum can occur when
 	\begin{eqnarray}
 		\gamma_{R}\gamma_{L}\Bigl|_\pm & = & \frac{1}{8}\Bigl(\eta^{2}\Omega^{2}+2\Gamma^{2}\pm2\eta\Omega\sqrt{\eta^{2}\Omega^{2}+2\Gamma^{2}}\Bigr).\label{eq:101}
 	\end{eqnarray}
 	Due to the constraint on \(\gamma_R\gamma_L\), i.e., \(0\leq\gamma_R\gamma_L\leq\beta^2\Gamma^2/4\), \(\gamma_R\gamma_L|_+\) in Eq.~\eqref{eq:101} can be ruled out. Since the other solution \(\gamma_R\gamma_L|_-\) does not always satisfy the lower bound of \(\gamma_R\gamma_L\), the condition under which \(\langle n_{1}\rangle_{\text{st}}\) can be minimized is
 	\begin{eqnarray}
 		2\Gamma^{2}&\geq&3\eta^{2}\Omega^{2}\label{eq:minGamma}
 		\\
 	 2\beta^{2}\Gamma^{2}&\geq&\eta^{2}\Omega^{2}+2\Gamma^{2}-2\eta\Omega\sqrt{\eta^{2}\Omega^{2}+2\Gamma^{2}}\label{eq:minBoundary}.
	\end{eqnarray}
	
Consequently, the best performance predicted in Eq.~\eqref{eq:100} still persists in the presence of nonguided decay if the total coupling efficiency \(\beta\) satisfies
	\begin{eqnarray}
	\beta\geq\beta_0 = \sqrt{1-\frac{\eta\Omega}{\Gamma^{2}}\Bigl(\sqrt{\eta^{2}\Omega^{2}+2\Gamma^{2}}-\frac{\eta\Omega}{2}\Bigr)}. \label{eq:betaLowerBound}
	\end{eqnarray}
	We choose four representative cases in Fig.~\ref{fig:Fig3} to show the emergence of \(\langle n_{1}\rangle_{\text{st}}^{\text{min}}\) at different \(\beta\), where Eq.~\eqref{eq:minGamma} is satisfied. The horizontal dashed and dotted line are the references of single ion cooling limit and the minimal phonon occupation predicted by Eq.~\eqref{eq:100}. For each \(\beta\in(\beta_0,1]\), we find that there are two \(\gamma_R^{\text{min}}\) corresponding to \(\langle n_{1}\rangle_{\text{st}}^{\text{min}}\), and they are located at
	\begin{equation}
		\gamma_{R}^{\text{min}}=\frac{1}{2}\beta\Gamma\pm\frac{1}{2}\sqrt{(\beta^{2}-1)\Gamma^{2}-\frac{\eta^{2}\Omega^{2}}{2}+\eta\Omega\sqrt{\eta^{2}\Omega^{2}+2\Gamma^{2}}}.
	\end{equation}
	In particular, the two \(\gamma_{R}^{\text{min}}\) are approaching \(\gamma_{R}=0.5\) as \(\beta\) decreasing from 1, and they coalesce at the point \(\beta= \beta_0\) which is shown in Fig.~\ref{fig:Fig3}(c). Once \(\beta<\beta_0\), as shown in Fig.~\ref{fig:Fig3}, the system no longer allows the optimal minimal \(\langle n_1 \rangle_{st}\) predicted by Eq.~\eqref{eq:100}, and the minimal \(\langle n_1 \rangle_{st}\) gradually regresses to single ion cooling limit.
	
\subsection{Superior cooling parameter regime}

	We now try to find the superior cooling parameter regime from Eq.~(\ref{eq:nstAnalytical}).
	It can be shown that $\langle n_{1}\rangle_{\text{st}}$
	exceeds $\langle n_{1}\rangle_{\text{st}}^{s}$ when
	\begin{equation}
		3\eta^{2}\Omega^{2}>2\Gamma^{2}-8\gamma_{R}\gamma_{L}\label{eq:WorseCond},
	\end{equation}
	and the superior cooling parameter regime~($\langle n_{1}\rangle_{\text{st}}<\langle n_{1}\rangle_{\text{st}}^{s}$)
	corresponds to
	\begin{equation}
		3\eta^{2}\Omega^{2}<2\Gamma^{2}-8\gamma_{R}\gamma_{L}\label{eq:SuperiorCooling}.
	\end{equation}
	We note that there is a constraint: $8\gamma_{R}\gamma_{L}\leq2\beta^{2}\Gamma^{2}$.
	This means that $\bigl\langle n_{1}\bigr\rangle_{\text{st}}$ can
	exceed $\bigl\langle n_{1}\bigr\rangle_{\text{st}}^{s}$ only when
	\begin{equation}
		\beta^{2}\geq1-\frac{3\eta^{2}\Omega^{2}}{2\Gamma^{2}},
	\end{equation}
	and the boundary of superior cooling parameter regime is determined by
	\begin{equation}
		\gamma_{R}^{s}=\frac{1}{2}\beta\Gamma\pm\frac{1}{2}\sqrt{(\beta^{2}-1)\Gamma^{2}+\frac{3}{2}\eta^{2}\Omega^{2}}.
	\end{equation}
	However, for the strong field regime such that \(\Gamma^2<3\eta^2\Omega^2/2\), every configurations of $\beta$
	and $\gamma_{R(L)}$ result in \(\langle n_{1}\rangle_{\text{st}}>\langle n_{1}\rangle_{\text{st}}^{s}\) according to Eq.~\eqref{eq:WorseCond}. Thus, we can only achieve superior cooling parameter regime when Eq.~\eqref{eq:minGamma} holds, under which \(\beta\) and \(\gamma_R\) can be tuned to realize the best performance in Eq.~\eqref{eq:100}.

\subsection{Cooling without nonguided decay (\texorpdfstring{$\beta=1$}{TEXT})}

	To discuss the chiral-coupling-assited cooling with the ideal chiral coupling~(\(\beta=1\)), we can adopt the result of Eq.~\eqref{eq:nstAnalytical} by setting \(\gamma=\Gamma=\gamma_R+\gamma_L\), which leads to Eq.~(5) in the main text
	 \begin{eqnarray}
	 	\bigl\langle n_{1}\bigr\rangle_{\text{st}} 
	 	& = & \frac{(\gamma_{R}-\gamma_{L})^{2}}{16\nu^{2}}+\Bigl(1+\frac{8\gamma_{R}\gamma_{L}}{\eta^{2}\Omega^{2}+2(\gamma_{R}-\gamma_{L})^{2}}\Bigr)\nonumber\\
		&&\times \frac{\eta^{2}\Omega^{2}}{8\nu^{2}}.\label{eq:57}
	 \end{eqnarray}
 
 	With the constraint $0\leq(\gamma_{R}-\gamma_{L})^{2}\leq\gamma^{2}$,
 	there are three values of $\gamma_{R}-\gamma_{L}$ that
 	determine the local extreme of $\langle n_{1}\rangle_{\text{st}}$:
 	\begin{eqnarray}
 		\gamma_{R}-\gamma_{L}\Bigl|_{\text{max}} & = & 0,\label{eq:58}\\
 		\gamma_{R}-\gamma_{L}\Bigl|_{\text{min}} & = & \pm\sqrt{-\frac{\eta^{2}\Omega^{2}}{2}+\eta\Omega\sqrt{\eta^{2}\Omega^{2}+2\gamma^{2}}}.\label{eq:59}
 	\end{eqnarray}
	Here, Eq.~(\ref{eq:58}) corresponds to the local maximum of $\bigl\langle n_{1}\bigr\rangle_{\text{st}}$:
 	\begin{equation}
 		\bigl\langle n_{1}\bigr\rangle_{\text{st}}^{\text{max}}=\frac{\gamma^{2}}{4\nu^{2}}+\frac{\eta^{2}\Omega^{2}}{8\nu^{2}},
 	\end{equation}
 	and Eq.~(\ref{eq:59}) corresponds to the same local minimum $\bigl\langle 	n_{1}\bigr\rangle_{\text{st}}$:
 	\begin{equation}
 		\bigl\langle n_{1}\bigr\rangle_{\text{st}}^{\text{min}}=\frac{\eta\Omega}{8\nu^{2}}\sqrt{\eta^{2}\Omega^{2}+2\gamma^{2}}-\frac{\eta^{2}\Omega^{2}}{32\nu^{2}}.
 	\end{equation}
 
	In addition, the superior cooling parameter regime~($\langle n_{1}\rangle_{\text{st}}<\langle n_{1}\rangle_{\text{st}}^{s}$) is given by Eq.~\eqref{eq:SuperiorCooling} at \(\gamma=\Gamma=\gamma_R+\gamma_L\),
	\begin{equation}
 		3\eta^{2}\Omega^{2}<2(\gamma_{R}-\gamma_{L})^{2},
	\end{equation} which is a straight line in $\Omega$-$\gamma_{R}$ plot as shown in Fig. \ref{Fig2}(b).

\end{document}